\documentclass[prl,twocolumn,groupedaddress]{revtex4}
\usepackage[dvips]{color}
\usepackage{graphicx,color}
\usepackage{bm}        
\usepackage{amssymb}   
\usepackage{amsmath}
\usepackage{cases}
\usepackage{mathrsfs}

\begin{document}

\title{Is 1T-TaS$_2$ a quantum spin liquid?}

\author{K. T. Law $^{1}$, Patrick A. Lee$^{2}$}\thanks{palee@mit.edu}

\affiliation{$^1$Department of Physics, Hong Kong University of Science and Technology, Clear Water Bay, Hong Kong, China}
\affiliation{$^2$ Department of Physics, Massachusetts Institute of Technology, Cambridge MA 02139, USA}

\begin{abstract}  \bf
1T-TaS$_2$ is unique among transition metal dichalcogenides in that it is understood to be a correlation driven insulator, where the unpaired electron in a 13 site cluster experiences enough correlation to form a Mott insulator. We argue based on existing data that this well-known material should be considered as a quantum spin liquid, either a fully gapped $Z2$ spin liquid or a Dirac spin liquid. We discuss the exotic states that emerge upon doping and propose further experimental probes.
\end{abstract}

\maketitle

\emph{Significance Statement} --- In solids with an odd number of electrons per unit cell, band theory requires that they are metals but strong interaction can turn them into insulators, called Mott insulators. In this case the electrons form local moments which in turn form an anti-ferromagnetic ground state. In 1973, P. W. Anderson proposed that in certain cases, quantum fluctuations may prevent magnetic order and result in a paramagnetic ground state called a quantum spin liquid. After many years of search, a few examples have been discovered in the past several years. We point out that a material which has been studied for many years, TaS$_2$, may be a spin liquid candidate. We point out further experiments which probe the exotic properties of this new state of matter.

\emph{Introduction}--- The transition-metal dichalcogenide (TMD) is an old subject that has enjoyed a revival recently due to the interests in their topological properties and unusual superconductivity. The layer structure is easy to cleave or intercalate and can exist in single layer form [\onlinecite{Geim, Klemm}]. These materials were studied intensively in the 1970Õs and 80Õs and were considered the prototypical examples of charge density waves (CDW) systems [\onlinecite{Wilson}]. Due to imperfect nesting in two dimensions, in most of these materials the onset of CDW gap out only part of the Fermi surface, leaving behind a metallic state which often becomes superconducting. Conventional band theory and electron-phonon coupling appear to account for the qualitative behavior [\onlinecite{Wilson}]. There is however one notable exception, namely 1T-TaS$_2$.  The Ta forms a triangular lattice, sandwiched between two triangular layers of S, forming an ABC type stacking. As a result the Ta is surrounded by S forming an approximate octahedron. In contrast the 2H-TaS$_2$ forms an ABA type stacking and the Ta is surrounded by S forming a trigonal prism. In a single layer, inversion symmetry is broken in 2H structure. The system is a good metal below the CDW onset around 90K and eventually the spin-orbit coupling gives rise to a special kind of superconductivity called Ising superconductivity [\onlinecite{Lu, Xi, Sato}]. In 1T-TaS$_2$ inversion symmetry is preserved. The system undergoes a CDW transition at about 350K with a jump in the resistivity. It is known that this transition is driven by an incommensurate triple-Q CDW (ICDW). A similar transition is seen in 1T-TaSe$_2$ at 470K. However, whereas TaSe$_2$ stays metallic below this transition, 1T-TaS$_2$ exhibits a further resistivity jump around 200K which is hysteretic, indicative of the first order nature of this transition. These transitions are also visible in the spin susceptibility data shown in Fig. 2. In early samples, the resistivity rises only by about a factor of 10 as the temperature is lowered from 200K to 2K and below that obeys Mott hopping law (log resistivity goes as $T^{-1/3}$) [\onlinecite{DiSalvo}]. More recent samples show better insulating behavior and it is generally agreed that the ground state is insulating. The 200K transition is accompanied by a lock-in to a commensurate CDW (CCDW), forming a $\sqrt{13} \times \sqrt{13}$ structure. As shown in Fig.1 this is described as clusters of Òstars of DavidÓ where the sites of the stars move inward towards the site in the middle. The stars of David are packed in such a way that they form a triangular lattice. Thus the unit cell is enlarged to have 13 Ta sites. The formal valence of Ta is 4+, and each Ta site has a single 5$d$ electron. We have an odd number of electron per unit cell. (We first restrict ourselves to a single layer.  Interlayer effects will be discussed later.) According to band theory, the ground state must be metallic. The only option for an insulating ground state in the pure material is a correlation driven Mott insulator. This fact was pointed out by Fazekas and Tosatti in 1976 [\onlinecite{Fazekas1}]. Band calculations show that band folding creates a cluster of bands near the Fermi surface. Rossnagel and Smith [\onlinecite{Rossnagel}] found that due to spin-orbit interaction a very narrow band is split off which crosses the Fermi level with a 0.1-0.2 eV gap to the other sub-bands. The narrow bandwidth means that a weak residual repulsion is sufficient to form a Mott insulator, thus supporting the proposal of Fazekas and Tosatti and distinguishes 1T-TaS$_2$ from other TMD. Apparently the formation of the commensurate clusters is essential for the strong correlation behavior in these 4$d$ and 5$d$ systems. Currently the assignment of Òcluster Mott insulatorÓ to the undoped 1T-TaS$_2$ ground state is widely accepted. A band about 0.2eV below the Fermi energy has been interpreted as the lower Hubbard band in ARPES [\onlinecite{Perfetti1,Ang}] and the electronic driven nature of the 200K transition has been confirmed by time dependent ARPES [\onlinecite{Perfetti2}].

\begin{figure}
\centering
\includegraphics[width=3in]{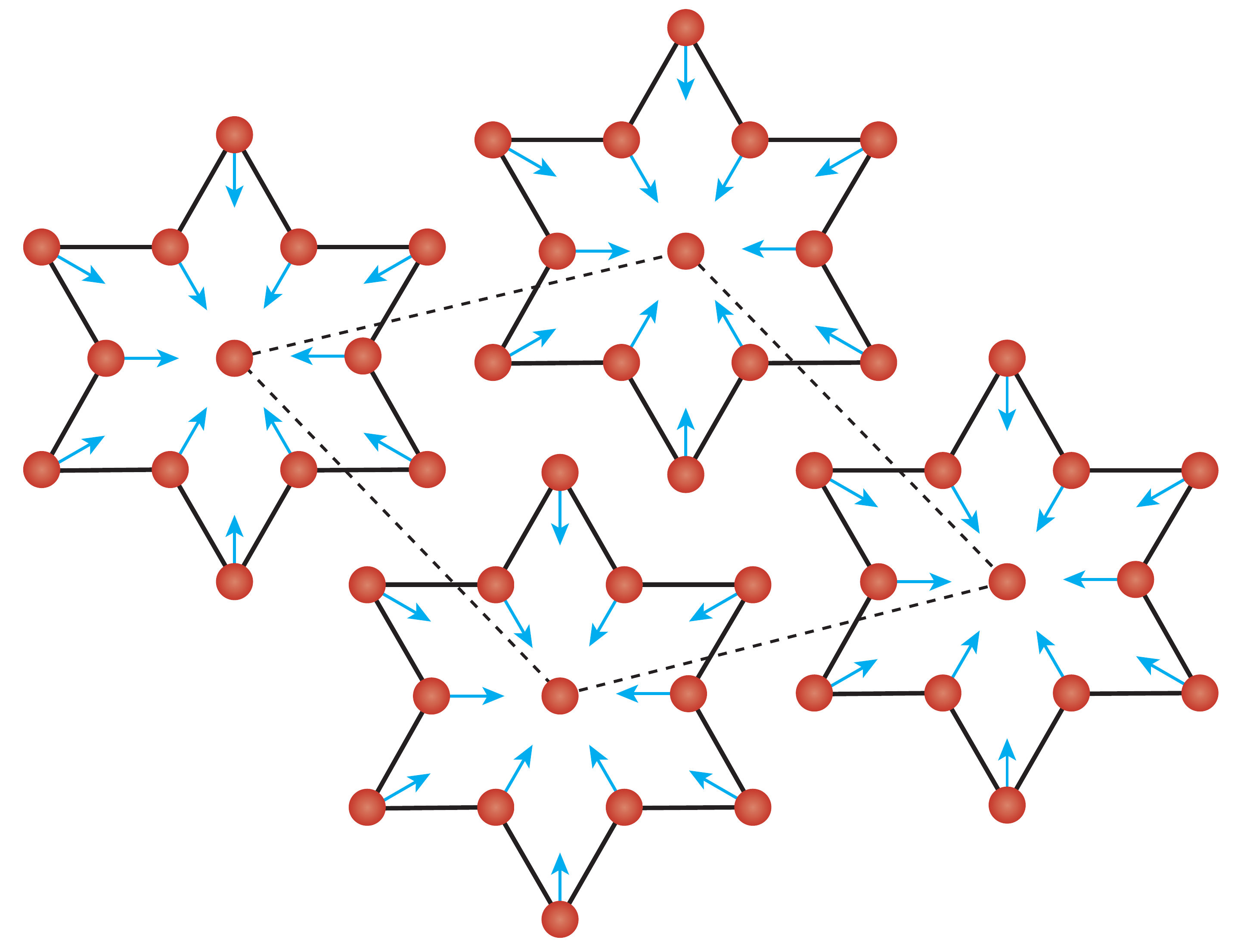}
\caption{In the cluster Mott phase of 1T-TaS$_2$, Ta atoms (red dots) belonging to a star of David move toward the Ta atom at the center. 13 Ta atoms form a unit cell and these unit cells form a triangular lattice. The directions and length of the arrows are schematic. }
\end{figure}

Surprisingly, with a few exceptions the issue of magnetism associated with the Mott insulator has not been discussed in the literature. In the standard picture of a Mott insulator, the spins form a local moment which then form an anti-ferromagnetic (AF) ground state due to exchange coupling. No such AF ordering has been reported in 1T-TaS$_2$.  There is not even any sign of the local moment formation, which usually appears as a rise in the spin susceptibility with decreasing temperature, following a Curie Weiss law. As seen in Fig.2, the magnetic susceptibility drops at the CDW transitions, but remains almost flat below 200K [\onlinecite{Wilson}]. (The small rise below 50K can be attributed to $5 \times 10^{-4}$ impurity spin per Ta). This serious discrepancy from the Mott picture did not escape the attention of Fazekas and Tosatti [\onlinecite{Fazekas1}]. They attempted to explain this by arguing that the g-factor is very small due to spin-orbit coupling. Their argument goes as follows. They assume cubic symmetry for the Ta $d$ state and that the lower state is $t_{2g}$, given by the basis $xy, xz,yz$. The 3 states form an effective $L=1$ basis set and with spin-orbit coupling, it splits into a $J=1/2$ doublet and $J=3/2$ quadruplet. They assume the quadruplet is lower and showed that the matrix elements of $L+2S$ vanishes in this manifold and there is no first order Zeeman effect. However this conclusion depends very sensitively on the orbital assignment. The Ta environment is far from cubic and the cluster wave-function depends on many details. In fact, Rossnagel and Smith [\onlinecite{Rossnagel}] claim that a single band crosses the Fermi level which is mainly made up of $xy$ and $x^2-y^2$ orbitals. This suggests that the low energy physics can be described by a single band Hubbard model. Due to the strong spin-orbit coupling, spin is not a good quantum number, but there is a pseudo-spin made up of Kramers pair to describe the degeneracy of each band in the presence of inversion symmetry. So throughout this paper, spin will refer to this pseudo-spin. In general  there is no $SU(2)$ symmetry for the pseudo-spin and we expect a non-zero and anisotropic g-factor. Thus we believe we cannot rely on the Fazekas-Tosatti proposal to explain the absence of Curie Weiss behavior.

A more recent discussion on magnetism came in 2005, when Perfetti et al. [\onlinecite{Perfetti1}] proposed  the existence of a fluctuating spin density wave (SDW) in order to explain certain features in their ARPES data. In the CCDW phase they reported a band about 0.15 eV below the Fermi energy which has a small upward dispersion. In addition, they reported spectral weight above this band, consistent with a back-folded band with a downward dispersion, which is greatly smeared. They compared this with the dispersion calculated for a triple-Q SDW and argued that their observation supports some kind of incipient SDW order. The spin density of this SDW is smoothly connected to that of the 3 sublattice 120 degree spin order of local moments expected for the Heisenberg model on a triangular lattice. Presumably the SDW fails to order due to quantum fluctuation. If so, their picture is smoothly connected to a spin liquid state formed out of local moments, even though there is no sign of local moments from the SDW picture.  It should be pointed out, however, that the more recent ARPES data while equally broad in energy, do not seem to show this downward dispersing feature [\onlinecite{Ang}]. 

In the past ten years there have been great advances in the study of spin liquids. (For review see [\onlinecite{Lee1}][\onlinecite{Balents}]). Originally Anderson [\onlinecite{Anderson}] and Fazekas and Anderson [\onlinecite{Fazekas2}] proposed a spin lquid state for the triangular lattice due to frustration. Now we have strong evidence to support the spin liquid state in the organic compounds which are close to the Mott transition as well as candidates in the Kagome lattice. In light of our recent experience, we can say that existing data are pointing strongly towards a quantum spin liquid state for the 1T-TaS$_2$ Mott insulator. Due to the rather low value of the sheet resistance, it is generally believed that disorder effects in the un-doped samples are small enough that we can rule out an insulator due to Anderson localization. There has been no report of further lattice distortion or phase transition below 200K, but in principle, long range order could set in immediately at 200K. However, a long range ordered spin density wave at wave-vector $Q$ will induce a charge density wave order at $2Q$, but these new diffraction spots have not been seen. Furthermore the ARPES should show clearly the back-folded band instead of a broad smear [\onlinecite{Perfetti1,Ang}].  Strictly speaking, we do not know of direct evidence that rules out static magnetization using techniques such as NMR or $\mu$SR. However, since the initial submission of this paper, we have learned that NMR and $\mu$SR data indeed rule out static magnetic moments.[\onlinecite{Klanjsek}],[\onlinecite{Kratoch}].  We therefore believe long range AF or SDW is unlikely. Instead 1T-TaS$_2$ may be an example of the elusive quantum spin liquid.

The purpose of this paper is to bring this exciting possibility to the attention of the community and discuss what kind of spin liquids are consistent with existing data. We also discuss further experiments which can be done to probe this new state of matter, but before that we need to discuss complications due to interlayer effects. Above 200K the ICDW are stacked in an ABC pattern, leading to a tripling of the unit cell in the c direction. However, below 200K the star of David are stacked directly on top of each other to form bi-layers. These bi-layers are stacked randomly or in an incommensurate fashion. The doubling of the unit in the bi-layer means that we now have an even number of electrons per unit cell. An obvious option is for the spins to form singlets between the unit cells. Strickly speaking, this state no longer fits the definition of a spin liquid, but is more analogous to ladder compounds. [\onlinecite{Rice}] Indeed a number of papers [\onlinecite{Durancet}] have suggested that the interlayer hopping dominates over the intra-layer hopping, and the system forms one dimensional metals. This point of view is supported by LDA plus $U$ calculations.[\onlinecite{Durancet}] However, we find this scenario unlikely, in view of the conclusion by Rossnagel and Smith [\onlinecite{Rossnagel}] that the split off bands are mainly $xy$ and $x^2-y^2$ orbitals   which, unlike the $z$ orbital, have weak interlayer overlap. In contrast Duracet et al [\onlinecite{Durancet}] first ignore spin-orbit coupling and split off an isolated band at the Fermi surface using the LDA+$U$ approximation. The $z$ orbital is strongly admixed and gives a strong inter-layer hopping. We believe it is more appropriate to first construct the best single particle orbital including spin-orbit coupling before turning on $U$ and that the conclusion of Duracet et al [\onlinecite{Durancet}] may be an artifact of the LDA plus $U$ approximation. There is some confirmation of the view that the inter-layer hopping is not dominant from the $1/T_1$ data of ref.[\onlinecite{Klanjsek}] because they do not find an exponential decay with temperature as the dominant inter-layer singlet model would imply.  Instead, they find a $T^2$ behavior which may support a Dirac type spin liquid in the temperature range between 50K and 200K. While the spin liquid may not be the true ground state for bulk crystals, it will be interesting to look at ultra-thin crystals. For example, the $\sqrt{13} \times \sqrt{13}$ structure in free-standing  tri-layer  crystals has been reported to be more robust than in the bulk, being stable even at room temperature. [\onlinecite{Sakabe}] It will also be interesting to grow atomically thin samples by MBE.

\emph{Possible Spin Liquid States in TaS$_2$} ---  Spin liquids can be divided into 2 broad classes, gapped or gapless. The gapless spin liquids generically has fermionic spinons which may form a Fermi surface or Dirac nodes [\onlinecite{Lee}]. The spinon Fermi surface is characterized by a linear term in the heat capacity with coefficient $\gamma$ given by a mass corresponding to a hopping matrix element of order $J$, the exchange energy [\onlinecite{Yamashita1}]. (Gauge fluctuations is predicted to convert the linear term to $T^{2/3}$ power, but it has proven difficult to distinguish between the two over the limited temperature range available in experiment.) A second key signature is a linear $T$ term in the thermal conductivity, which is usually observed only in a metal [\onlinecite{Yamashita2}]. It is widely believed that the spin liquid observed in the organics belong to this type [\onlinecite{Lee1,Balents}]. Interestingly, the heat capacity of 1T-TaS$_2$ shows a linear intercept with small upturn at low temperature [\onlinecite{Benda}]. The coefficient of this linear term is about 2mJ/Mole-$K^2$, about 4 times that of Copper. This corresponds to a Fermi energy of about 0.16 eV. We note that the bandwidth of the band at the Fermi level from band theory is considerably narrower than this [\onlinecite{Rossnagel}]. Since we expect $J$ to be at least several times smaller than the bandwidth, the observed $\gamma$ is much too small to be due to a spinon Fermi surface. It is probably due to the impurity moments seen in the spin susceptibility, forming a random singlet type state. (In the organics the $\gamma$ is several tens of mJ/Mole-$K^2$ [\onlinecite{Yamashita1}] and corresponds to $J$ of 250K). Thus the specific data effectively rule out a spinon Fermi surface. A Dirac spin liquid remains a possibility.

Next we come to the gapped spin liquids. In mean-field theory, the most common description is one with gapped bosonic spinons [\onlinecite{Read}] together with gapped visons [\onlinecite{Senthil1}]. These are called $Z2$ spin liquids. We note that in two dimensions, $U(1)$ gapped spin liquid is not allowed, because the compact $U(1)$ gauge field will lead to confinement. A fermionic mean field theory can also lead to a $Z2$ state with fermionic spinons [\onlinecite{Baskaran}]. However, this state is smoothly connected to the one with bosonic spinons, because a bosonic spinon can bind with a vison to form a fermionic spinon, and it is the question of energetics as to whether the low energy spinons are fermions or bosons. There are more exotic possibilities, but our expectation is that without further breaking of symmetry, the $Z2$ gapped spin liquid is likely the only common possibility. This state is characterized by low energy excitations which are gapped spinons (fermions or bosons) and visons. To explain the spin susceptibility data, we will have to argue that the background subtraction in Fig.2 is such that the susceptibility has dropped to a  small value below the first order transition at 200K. This means that the spin gap and the exchange scale J is above 200K. This will allow us to get around the issue of the absence of the Curie-Weiss law due to local moments.


\begin{figure}
\centering
\includegraphics[width=3.2in]{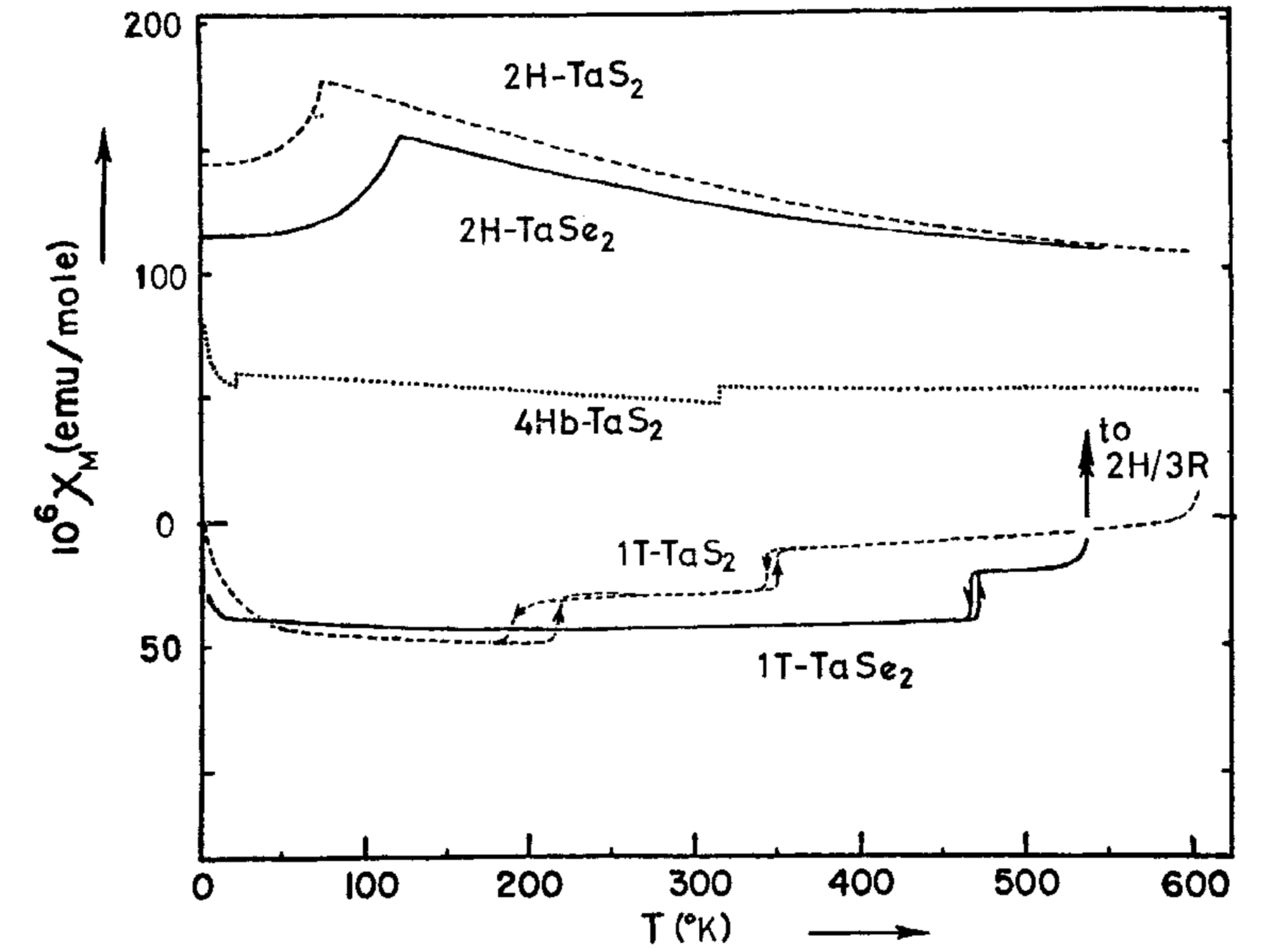}
\caption{The molar magnetic susceptibility ( $\chi_M$) versus temperature (T) for TaS$_2$ and T$_c$ with different lattice structures. The background diamagnetic term has not been subtracted. The data is taken from Ref.[\onlinecite{Wilson}]. }
\end{figure}

There is one more possibility of a gapped spin liquid that has been widely discussed for triangular lattice, and that is the chiral spin liquid [\onlinecite{Kalmeyer}]. This state spontaneously breaks time-reversal symmetry. There should be an easily detected Ising type transition at a finite temperature, but one could argue that it is not accessible because its temperature scale is above the first order transition at 200K. The chiral spin liquid has many observable consequences that has been widely discussed, including Kerr rotation, chiral spin edge modes, spontaneous quantized thermal Hall conductivity etc. The absence of signatures of time-reversal symmetry breaking makes this unlikely.

We note that recent numerical work using DMRG and variational Monte Carlo methods support the existence of a region of spin liquid in a $J_1-J_2$ Heisenberg model on a triangular lattice where the NNN $J_2$ is in the range $0.08<J_{2}/J_{1}<0.15$ [\onlinecite{Cenke,Zhu,Hu,Iqbal}]. Currently, many groups find it difficult to distinguish between the $Z2$ spin liquid, the chiral spin liquid and the $U(1)$ Dirac spin liquid. These states seem to have very similar energies. It is also important to remember that our system has strong spin-orbit coupling. Consequently there is no $SU(2)$ symmetry for the pseudo-spin and the Heisenberg model is not a good starting model. We expect anisotropic exchange terms and ring exchange terms. Thus the phase space may be quite large to support some form of spin liquid.

\emph{Effect of doping, pressure and further experimental consequences}--- Next we discuss what is known experimentally when the system is doped or when pressure is applied. It has been reported that 1\% Fe doping destroys the CCDW state [\onlinecite{Li}]. A Fermi surface appears near the $\Gamma$ point above this doping level [\onlinecite{Ang}] and superconductivity with about 3K T$_c$ appears. Further doping creates Anderson localization and kills the superconductivity. With a small amount of pressure of 1 GPa, the CCDW is destroyed and replaced by the ICDW, which is in turn destroyed with a pressure of 7 GPa [\onlinecite{Sipos}]. The state is metallic and superconducting with T$_c$ about 5K everywhere above 4 GPa. We summarize the low temperature state in the pressure and doping concentration plane in Fig 3.


\begin{figure}
\centering
\includegraphics[width=3.2in]{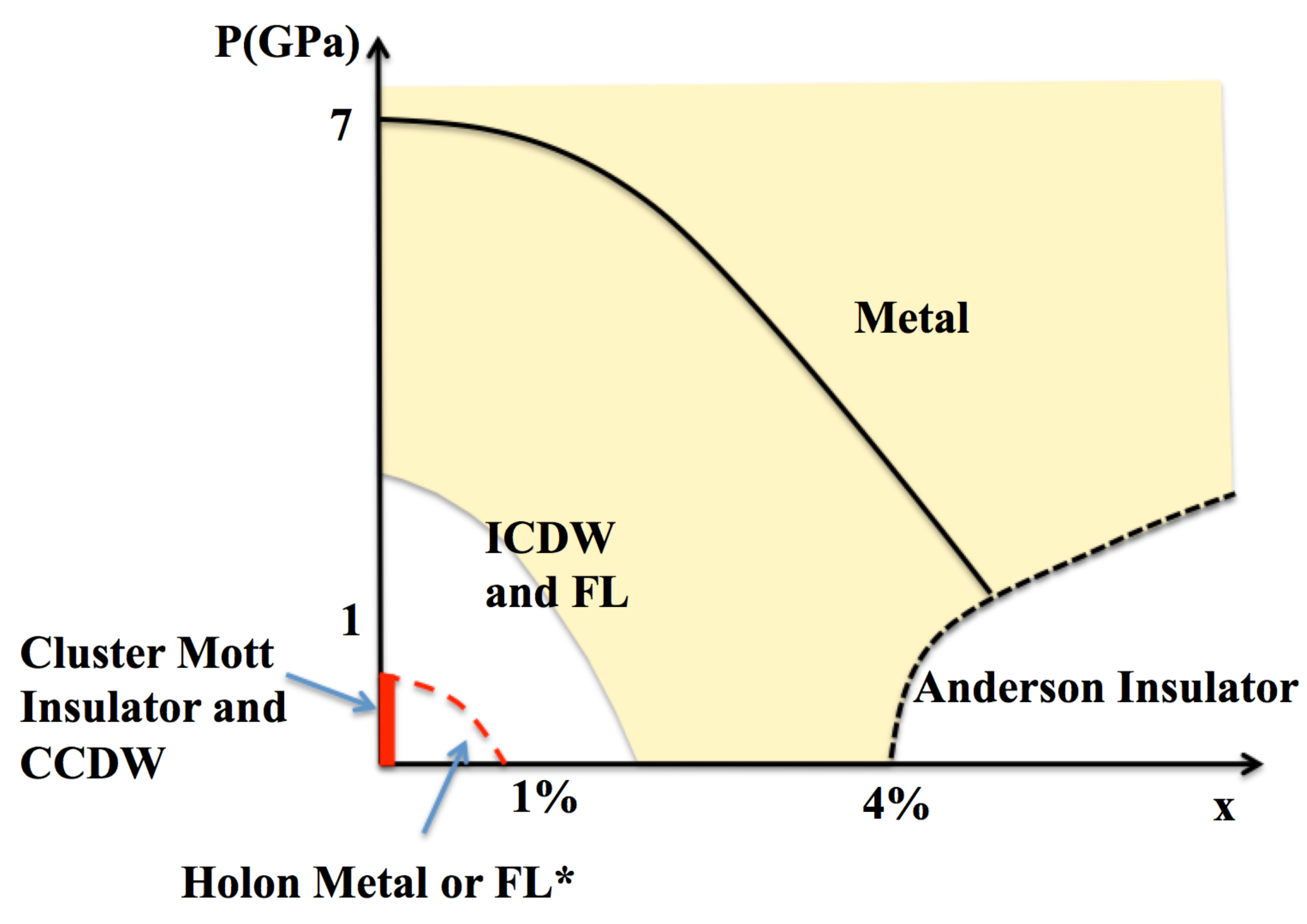}
\caption{ Schematic phase diagram of the low temperature state in the pressure $P$ and the doping concentration ($x$ per Ta) plane. Red dashed line marks a first order phase transition separating a commensurate charge density wave (CCDW) and a incommensurate charge density wave (ICDW). The solid red line marks the undoped spin liquid state formed out of the cluster Mott insulator, the cluster being the star of David shown in Fig.1. The solid black line separates the ICDW phase and a normal metal phase. The ICDW phase is a Fermi liquid (FL) metal satisfying the Luttinger volume of approximately $1/13+x$ of the original Brillouine zone volume. Dotted black line marks the onset of Anderson localization due to dopant disorder and is highly schematic. The yellow region denotes superconducting ground state with T$_c$ of approximately 3-5K. }
\end{figure}

It is worth noting that 1\% Fe doping is actually a rather large doping in our effective Hubbard model of clusters, because it corresponds to 13\% doping per cluster. Thus it is not surprising that the state looks like a conventional Fermi liquid, with a small and round Fermi surface near the $\Gamma$ point that presumably obeys the Luttinger volume [\onlinecite{Ang}]. (Strictly speaking there is no Luttinger theorem for an incommensurate state, but we can use the nearby commensurate approximate of $\sqrt{13} \times \sqrt{13}$ state to estimate the Fermi surface volume.) The observed superconductivity seems to grow out of this state and extend beyond a pressure where the ICDW is destroyed. Thus we think this superconductivity may be quite conventional. From the point of view of seeking exotic physics, the most interesting region is the small lower left corner in Fig.3 where the CCDW order is present. Unfortunately we are not aware of doping data below 1\% carrier per Ta. This region of low dopant density is prone to Anderson localization. Thus doping by substitution in the plane likely to create too much disorder. It will be good to attempt doping by intercalation or even better, by gate in a thin sample. (Focussing on mono-layer or tri-layer will also help us get around the issue of inter-layer coupling.) Indeed, limited data in the very lightly doping range showing metallic like behavior have reported by gate doping [\onlinecite{Yu}]. Here we describe the possible states that result from doping without disorder, and discuss some experimental consequences. Soon after the introduction of the concept on spinons [\onlinecite{Anderson2}] which carry spin and no charge, Kivelson, Rohksar and Sethna [\onlinecite{Kivelson2}]  pointed out that an analogous situation may obtain for the doped holes, which carry charge and no spin.  These are called holons. For very small doping, a natural state is a Wigner crystal of holons. [\onlinecite{Kivelson1}] Here we focus on states that do not break translation symmetry which may emerge with sufficient doping.

1.	Doping a $Z2$ gapped spin liquid. In the mean-field description with bosonic spinons, the simplest case is that fermionic holons form a Fermi surface with an area corresponding to $y$, which is defined as the density of dopant per cluster. ($y$=13$x$ if $x$ is dopant per Ta). For bi-partite lattices, there is an additional quantum number corresponding to the A, B sublattice occupied predominantly by the holons [\onlinecite{Shraiman}]. For the triangular lattice, this is not the case and the volume of the Fermi surface corresponds to $y$ spinless fermions. This state has been called the holon metal.

A second possibility is that a holon binds with a spinon to form a physical hole, which in turn forms a Fermi surface. Since the hole carries spin, the volume is smaller than the holon metal case by a factor of 2. This state has been called FL* because it behave like a Fermi surface, but does not obey the conventional Luttinger theorem with $1+y$ electron per cluster [\onlinecite{Senthil2}].

A third possibility is that the state is a superconductor. In mean field theory this emerges most clearly when the spinons are described as fermions. In this case the holons are bosons which will condense and form a conventional superconductor. This is the RVB route to superconductivity as envisioned by Anderson [\onlinecite{Anderson2}].

The holon metal and the FL* are clearly exotic metals because they are metallic ground state which dramatically violate Luttinger theorem. (Fermi surface volume of $y$ or $y/2$ versus $1+y$)  The physical hole or electron excitation is gapped in the holon metal (because the spinon is gapped) but is gapless in FL*. This distinction will show up clearly in ARPES and tunneling. The FL* will show a Fermi surface but the holon metal will appear as gapped. Nevertheless, in a clean enough sample, the holon metal will exhibit quantum oscillations. In a multi-layer systems, the stacked holon metal is insulating in the direction perpendicular to the layer (because only a physical hole can hop between planes) while the FL* state is metallic. As discussed earlier the two states can be distinguished by the size of the Fermi momentum, $k_F$, which can in principle be measured via Koln anomaly or via Friedel oscillations by STM imaging [\onlinecite{Mross}].

2.	Doping the  Dirac spin liquid. The Dirac spin liquid can be $Z2$ or $ U(1)$. The former will have gapped visons while the latter has dissipative gapless Ò gauge photonsÓ. A natural consequence of doping the Dirac spin liquid is that the bosonic holons condense, resulting in a nodal superconductor. Alternatively, the holons can bind with some of the spinons and form a Fermi surface of volume $y/2$, just like the FL* phase  in case 1. The left over spinons may form their own spinon Fermi surface.

3.	Doping a chiral spin liquid. The natural consequence is a gapped superconductor which breaks time-reversal symmetry. This has been much discussed, but the effect of spin-orbit coupling has not been explored. The FL* is also a possibility.

Finally, if the ground state is an inter-layer singlet, this situation is analogous to that in the ladder compounds in the context of cuprates, and it is possible that doping will induce inter-layer pairing of the doped holes, leading to superconductiviy.[\onlinecite{Rice}]

Is there any unique signature  of the spin liquid itself? It  has been proposed  that the undoped $Z2$ gapped spin liquid may show dramatic phenomena when placed in contact with superconductors or magnets.[\onlinecite{Senthil3}],[\onlinecite{Barkesh}] For example Senthil and Fisher [\onlinecite{Senthil3}] pointed out that if the spin liquid Mott insulator is used as the insulator barrier between two conventional superconductors in an S-I-S structure, this structure may exhibit a charge e Josephson effect. The requirement is that the spatial transition between the superconductor and the spin liquid has to be smooth enough to avoid confinement at the interface. Ordinarily this is a tall order. However the 1T-TaS$_2$ offers a special opportunity in that the superconductors can be created by doping. This can be achieved by gating, which can produce a smooth transition between the superconducting state and the insulating state.

In summary, we believe existing experimental data strongly point to 1T-TaS$_2$ as an  example of a spin liquid state, formed out of a cluster Mott insulator. The lightly doped state is likely an exotic metal with unusual experimental consequences, some of which we discussed. 

\section{Acknowledgement}
We thank Debanjan Chowdhury for bringing several key experimental papers to our attention and to him and T. Senthil for discussions and review of theoretical ideas. PAL acknowledges the NSF for support under grant DMR-1522575. KTL acknowledges the support of RGC and Croucher Foundation through HKUST3/CRF/13G, C6026-16W, 16324216 and Croucher Innovation Grant.

\end{document}